\documentclass[preprint,proceedings]{rmaa}

\def\spose#1{\hbox to 0pt{#1\hss}}
\def\lta{\mathrel{\spose{\lower 3pt\hbox{$\mathchar"218$}}
        \raise 2.0pt\hbox{$\mathchar"13C$}}}
\def\gta{\mathrel{\spose{\lower 3pt\hbox{$\mathchar"218$}}
        \raise 2.0pt\hbox{$\mathchar"13E$}}}

\def\msol{\rm M_\odot}

\SetYear{2003} \SetConfTitle{Compact Binaries in Galaxy and
beyond}

\title{The implications of disc instabilities on cataclysmic variable
structure and evolution }

\author{
  J.-P. Lasota\altaffilmark{1}}

\altaffiltext{1}{Institut d'Astrophysique de Paris, France}

\shortauthor{Lasota}
\shorttitle{RevMexAA(SC) Disc instabilities and truncation}

\fulladdresses{
\item J.-P. Lasota: Institut d'Astrophysique de Paris,
  98bis Bd Arago, 75014 Paris, France
  (\email{lasota@iap.fr}).
}

\listofauthors{J.-P. Lasota}
\indexauthor{J.-P. Lasota}

\abstract{Applications of the thermal-viscous disc instability
model to various classes of cataclysmic variable very often
require the accretion disc to be truncated. I argue that in most
cases this inner truncation must be due to the white dwarf's
magnetic field.}

\resumen{Spanish abstract - to be supplied by the editors.}

\addkeyword{Accretion discs}
\addkeyword{Instabilities}
\addkeyword{Stars: Cataclysmic}
\addkeyword{Stars: Evolution}

\begin{document}
\maketitle

\section{The magnetic nature of VY Scl stars}
\label{sec:vy}

VY Scl stars are very bright Cataclysmic Variables (CVs) which
occasionally undergo a diminution in brightness of more than one
magnitude. Such drops in luminosity should bring the accretion
discs in these systems into the dwarf nova instability strip, but
surprisingly no outbursts are observed. As pointed out by Lasota,
Hameury \& Hur\'e (1995) outbursts would be prevented if the disc
were truncated at radius
\begin{equation}
r_{\rm tr} > r_{\rm crit} \approx 6 \times 10^{9} \left(\frac{\dot
M}{10^{15}{\rm g\ s^{- 1}}}\right)^{0.375} M_1^{0.333}\; {\rm cm},
\label{eq:rin}
\end{equation}
where $r_{\rm tr}$ is the transition radius and $M_1$ is the
white-dwarf mass in solar units (see e.g. Lasota 2001). Such
truncation removes the inner, unstable parts of the disc. This
stabilization can be achieved either by the action of the white
dwarf's magnetic field or by heating and/or evaporation of the
unstable inner regions. Leach et al. (1999; see also King 1997;
Hameury, Lasota \& Dubus 1999) proposed that heating by the
accretion-heated white dwarf in highly accreting VY Scl stars
brings their inner disc into a hot and stable state. In such a
case the transition radius would be (Hameury \& Lasota 2002,
hereafter HL02):
\begin{equation}
r_{\rm tr} \approx 6.7 \; T_{*,40}^{4/3} \left(\frac{T_{\rm
crit}}{6500\ K}\right)^{-4/3} R_* \ ,
\label{eq:rtr}
\end{equation}
where it is assumed for simplicity's sake that a disc is thermally
stable if its photospheric temperature is greater than $T_{\rm
crit} \simeq 6500$ K, $T_{*,40}= T_*/40,000 K$ and the albedo=0
(which obviously maximizes the transition radius). Therefore a 0.7
$\msol$ white dwarf would stabilize the disc only if heated to
more than 50,000 K and more massive stars would have to be heated
to temperatures that have never been observed. In any case, RU
Peg, the CV with the hottest (50,000 K) white dwarf observed (Sion
\& Urban 2002), is a dwarf nova, so at least in this case
white-dwarf heating does not work.

In spite of these arguments and observations Hoard et al. (2003)
assert that the VY Scl star DW UMa could be stabilized by white
dwarf heating because it has the right temperature (50,000 K) and
its radius, due to heating, could be much larger than one would
deduce from its mass (0.8 $\msol$). Unfortunately the argument
cited in favour of oversized white dwarfs (Koester \& Schoenberner
1986) does not apply to the old primaries in CVs. Fortunately,
however, the case of DW UMa provides a very strong argument in
favor of a truncation that cannot be achieved by any existing hot
white dwarf. As argued by HL02 the real problem is not so much
with explaining the absence of outbursts during low states as
explaining their absence during long intermediate states. During a
very long transition from a low to a high state DW UMa showed no
outbursts. The 4 mag rise lasted $\sim 500$ days (Honeycutt et al.
1993), therefore the disc would have crossed the instability strip
in a quasi-stationary way, because its viscous time is only $\sim$
10 - 20 days. This is illustrated by Fig.~1 from HL02, which shows
the light-curve of a system in which the disc truncation prevents
outbursts during quiescence, while a frantic outburst activity is
observed during intermediate states. The {\sl only} thing that
would prevent outbursts during long intermediate states would be
the absence of the disc, i.e. the transition radius should be at
least of the order of the circularization radius, say. Therefore
despite Hellier's (this volume) comment that ``... Hameury \&
Lasota (2002) prefer a model in which a strong magnetic field
evacuates the inner disc", the question is not of preference,
strong field or inner disc. In order for the inner disc to be
evacuated there must be typically magnetic moments $\mu=5\times
10^{30}$ G~cm$^3$, therefore rather weakish magnetic fields. But
the absence of outbursts during intermediate states requires the
evacuation of the ``whole" disc, and this does indeed require
stronger fields: $\mu \gta 4 - 7.5 \times 10^{32} {\; \rm G\
cm^3}$, i.e. magnetic moments expected in some Intermediate Polars
(IP).
\begin{figure}[t!]
  \includegraphics[width=\columnwidth]{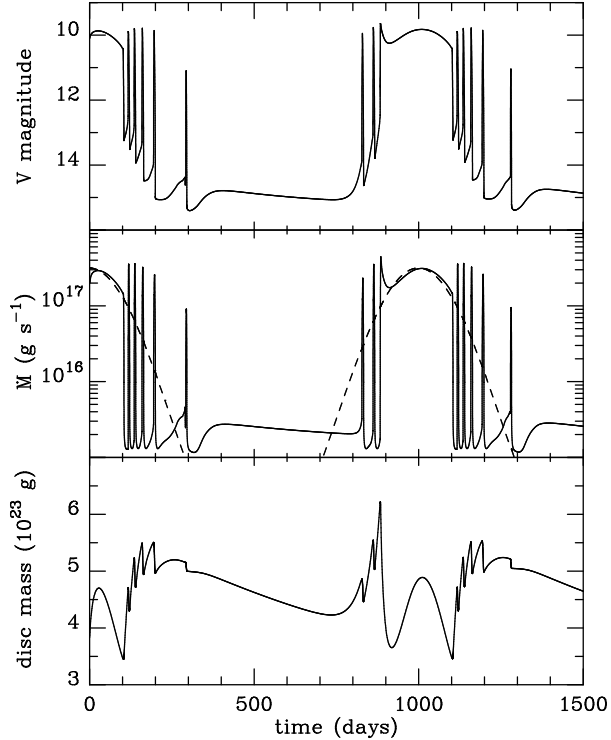}
  \caption{
Top panel : visual light curve of a binary system in which the
mass-transfer rate slowly varies. The accretion disc is disrupted
by the magnetic field of a 0.7 $M_{\odot}$ white dwarf with
$\mu_{30}$= 10. Intermediate panel : mass accretion rate onto the
white dwarf (solid line), and mass transfer rate (dashed line).
Bottom panel : disc mass}
  \label{fig:trunc}
\end{figure}

This would imply that variable circular polarization should be
detected, at least in such VY Scl stars as DW UMa or MV Lyr in
which long intermediate states are observed. One should keep in
mind, however, that circular polarization has been detected only
in five IPs out of a total of more than 30 -- those that harbor
white dwarfs with the highest magnetic fields $\sim 2 - 8 \rm~MG$,
the highest value corresponding to V2400 Oph (Buckley et al.
1995). Since VY Scl stars would have rather lower fields  , the
detection of their circular polarization could be rather
difficult.

Another signature of the presence of a magnetized white dwarf
takes the form of periodic oscillations and none seem to be
detected so far in VY Scl stars. But such detection could be
difficult, as shown by the case of one of the best-observed
cataclysmic variables, the SU UMa type dwarf nova OY Car.
Indications of the presence of a magnetic white dwarf in this
short period ($P_{\rm orb}=1.51$ h), eclipsing ($i=83^o$) and not
very distant ($\sim 82$~pc) have been found only recently (Ramsay
et al. 2001a; Hakala \& Ramsay 2003; Wheatley \& West 2003). The
evidence comes from X-ray observations of OY Car in quiescence.
Considering that it took more than twenty years to find hints of
the magnetic nature of this so frequently observed system and that
the evidence is still not conclusive it should not be surprising
that observations of VY Scl stars have not yet provided such
evidence. The nature of their X-ray emission is subject to
controversy which could be resolved by Chandra and XMM grating
spectra observations (Mauche \& Mukai 2002).

\section{Magnetic dwarf novae}
\label{sec:dn}

If magnetic, OY Car would have  a truncated disc in quiescence. It
is too often forgotten that in quiescence dwarf-nova accretion
discs are {\sl unsteady}, i.e. the accretion rate through the disc
is {\sl not} constant. According to the Disc Instability Model
(DIM) the quiescent accretion rate in the disc $\dot M_{\rm q}(r)$
must be everywhere smaller than the critical rate $\dot{M}_{\rm
cr}$:
\begin{equation}
  \label{eq:one}
\dot M_{\rm q} < \dot{M}_{\rm cr} = 9~\times ~ 10^{12} ~\left(
{M_1 \over \rm M_\odot} \right)^{-0.88} r_9^{2.65} ~\rm g~s^{-1}
\end{equation}
where $r_9$ is the radius in units of $10^{9}$ cm. Therefore in a
quiescent disc extending down to the surface of the accreting star
the inner accretion rate would be very low, in general more than
two orders of magnitude lower than the rate at which matter is
transferred from the companion to the outer disc regions. But
X-ray observations of quiescent dwarf novae often imply accretion
rates higher than the critical rate near the white dwarf's
surface. The accretion rate producing the observed X-ray
luminosity $L_X$ can be estimated as
\begin{equation}
  \label{eq:xraymdot}
\dot{M}_{\rm X} = 7.5~\times ~ 10^{13}\eta^{-1}_X r_9~\left( {M_1
\over \rm M_\odot} \right)^{-1} \left(\frac{L_X}{10^{31} {\rm
erg~s^{-1}}}\right)~\rm g~s^{-1}
\end{equation}
where $\eta_X$ is the efficiency of producing X-rays. From the
estimated bolometric X-ray luminosity $L_X \sim 4 \times 10^{30}
\rm erg~s^{-1}$ of OY Car in quiescence, Ramsey et al. (2001b) one
can deduce, for a 1 $M_{\odot}$ white dwarf, an accretion rate
$\dot M_{\rm X} \approx ~1.5 \times 10^{13}\eta^{-1}_X~ \rm
g~s^{-1}$, whereas the critical rate near the white dwarf's
surface is $\sim 1.4 \times 10^{12}~\rm g~s^{-1}$, i.e. at least
($\eta_X < 1$) ten times lower. Taken at face value this would
mean that the inner disc in OY Car when it is quiescent is
unstable. Clearly it is not. As pointed out by Meyer \&
Meyer-Hofmeister (1994) a truncated disc would remove this
inconsistency. (These authors argue in favour of disc truncation
by evaporation.) This would require the inner disc radius to be
\begin{equation}
  \label{eq:xrayr}
r_{\rm in} > 9.6 \times 10^{9}~
\eta^{-0.61}_X~M_1^{0.67}\left(\frac{L_X}{10^{31} {\rm
erg~s^{-1}}}\right)^{0.61} {\rm cm},
\end{equation}
which for OY Car would require $\mu \gta 8 \times 10^{30} \rm
G~cm^{3}$.

This is slightly higher than the magnetic moment ($2 \times
10^{30} \rm G~cm^{3}$) producing the truncation required to
reproduce the observed quiescent X-ray luminosity \underline{and}
the time-lags observed in the X-ray, EUV, UV and optical emission
of SS Cyg (Schreiber, Hameury \& Lasota 2003).

\section{WZ Sge as a magnetic rotator}
\label{sec:wz}

The WZ Sge 27.87 s coherent oscillation was first attributed to
the rotation of an accreting magnetized white dwarf by Patterson
(1980). This oscillation disappeared during the 1978 outburst and
was absent for about 16 years (although a 28.96 s oscillation
reappeared when WZ Sge was at twice its pre-outburst brightness).
Although it was easy to see that during the outburst the increased
accretion rate could suppress the magnetosphere, the lack of the
principal pulse afterwards was harder to understand and cast a
shadow of doubt on the presence of a rapidly rotating white dwarf
in WZ Sge. However, in 1995 the 28 s oscillation reappeared and
was detected also in the 2-6 keV ASCA energy band (Patterson et
al. 1998). This strengthened the case for  the presence of a
rotating, magnetized white dwarf in WZ Sge, especially since UV
spectral observations (Cheng et al. 1997) confirmed the presence
of a rapidly rotating white dwarf. Then, WZ Sge unexpectedly
($\sim 7$ years early) went to outburst in 2001 and once again the
28 s signal was lost. Observations during the decline and after
the system settled back to quiescence led to two rather confusing
conclusions. First, the 29 s oscillation, which reappeared early
as during the 1978 outburst, was shown to be independent of the
white dwarf's temperature (Welsh et al. 2003), thus ruling out the
the white dwarf nonradial g-mode pulsation hypothesis which had
proposed as an alternative to positing a rotational origin to the
oscillations. Interestingly, such pulsations have been recently
detected in GW Lib (van Zyl et al. 2004) a ``near twin of WZ Sge"
(Thorstensen et al. 2002).

On the other hand the 27.87s is again absent, this time also in
XMM-Newton data which obviously challenges the magnetic rotator
model of WZ Sge (Mukai, these proceedings). Of course one should
remember that after the 1978 outburst there was a wait of 16 years
before this signal reappeared. So maybe it was risky for the
present author to bet a bottle of ``Paradigm" wine that the 28 s
would reappear after only 7 years after the last outburst. The
hope lies in the high sensitivity of XMM-Newton.

In any case one has to agree with Koji Mukai that ``if it [WZ Sge]
is magnetic it does not resemble any other magnetic binary we know
about. This in fact was the verdict of  Lasota, Kuulkers \&
Charles (1999), who concluded that WZ Sge is in an `ejector'
phase, ejecting most of the transferred matter, in a state
``totally different from the usual IPs". WZ Sge is most probably
in a state intermediate between that of AE Aqr -- a pure discless
ejector (Wynn et al., 1997) and that of DQ Her which seems to have
a steady accretion disc. No model of such a flow has been yet
published but one should appear soon (Wynn, private
communication).

If the white dwarf in WZ Sge is magnetic we have to try to
understand how it has been spun up to its present rate. The only
way to achieve this is through accretion disc spin-up. The white
dwarf spin rate increases as it accretes the Keplerian angular
momentum of matter at the inner disc edge. An equilibrium period
is reached when angular momentum is accreted at the same rate as
it is centrifugally expelled by the spinning white dwarf. In the
case of a magnetized white dwarf the equilibrium period is equal
to the Keplerian period of the magnetosphere ($\omega_s=1$). The
equilibrium spin would be then
\begin{equation}
P_{\rm eq} = 360 M_1^{-5/7} \dot M_{15}^{-3/7} \mu_{31}^{6/7} \
{\rm s} \label{peq}
\end{equation}

From Eq. (5) and the quiescent X-ray luminosity $L_X \sim 3 \times
10^{30} \rm erg~s^{-1}$ one can deduce a magnetic moment $\sim
10^{31} \rm G~cm^{3}$. According to the ejector model of Lasota,
Kuulkers \& Lasota (1999) the magnetic moment could be even
greater. This means that an accretion rate $\gta 10^{17}$ g
s$^{-1}$ was required to bring WZ Sge to its present very rapid
rotation rate. Such accretion rates are typical of nova-like CVs
at orbital periods longer than 3 hours whereas WZ Sge is very
close to the minimum period for CVs where secular mass-transfer
rates are two orders of magnitude lower. The present mass transfer
rate in WZ Sge is in good agreement with that predicted by mass
loss from the secondary driven by gravitational radiation alone.
However, fluctuations of the mass-transfer rate on time-scales of
the order of the white dwarf spin-up time, i.e. 10$^4-10^5$ years,
would not modify the secular evolution of the binary.

The hypothesis of high mass-transfer rate episodes in the life of
WZ Sge is supported by the existence of ER UMa stars. These
systems have orbital periods between 79 and 92 minutes and should
have high ($\sim 10^{17}$ g s$^{-1}$) mass transfer rates, which
would explain their extremely short intervals between
superoutbursts (19 -- 44 d) and very short (3 - 4 d) `normal'
outburst intervals. They are the high accretion-rate equivalent of
the low accretion rate systems SU UMa's (OY Car is such a system,
WZ Sge is a SU UMa system showing superoutbursts only). The idea
that DI UMa (one of the 4 known ER UMa's with $P_{\rm orb} =79$
min), which is more luminous than WZ Sge by a factor $\sim 50$,
spends most of its life as `an ordinary WZ Sge star', but was
caught during an `upward surge in accretion', has been proposed by
Patterson (1998).

The white dwarf in WZ Sge was spinning at the equilibrium rate
when this system was an ER UMa star.  After some time the mass
transfer rate would return to its secular value, two orders of
magnitude lower than that required for $P_{\rm eq}\sim {\rm few}
\times 10$ s.  In a very short time (the viscous time of the disc
$\sim$days) the magnetosphere will start expanding and becoming
larger than the corotation radius. The system enters into the
ejector phase.  It will stay in this, WZ Sge-phase (about $10^5$
years) until it gets to a new spin equilibrium corresponding to
the low accretion rate ($\sim$ a few minutes). HT Cam is a CV with
exactly such properties: $P_{\rm orb}=86$ min, $P_{\rm spin}=8.6$
min. The spin period corresponds to the equilibrium value as given
by Eq. (6) for the expected secular mass-transfer rate.  One can
therefore expect  the white dwarf's magnetic field in HT Cam to
have a strength close to that of WZ Sge. and this system would be
a WZ Sge-type system at a particular phase of its
spin-up/spin-down history.

A jump in the mass transfer rate could bring such an equilibrium
system back to a spin-up phase. If HT Cam does indeed have a
magnetic field close to that of WZ Sge and if its mass-transfer
rate is close the secular rate, its accretion disc should be
truncated at a radius close to the value given by Eq. (1) and
therefore stable with respect to the thermal-viscous instability.
HT Cam shows very brief outbursts (Ishioka et al. 2002; Kemp et
al. 2002) that could reflect the marginal stability of its disc.

\section{Conclusion}

There is growing evidence that accretion discs in CVs are
truncated. Although I prefer to posit a magnetic origin for this
effect, I cannot preclude that in some cases it is due to disc
evaporation as proposed by Meyer \& Meyer-Hofmeister (1994),
especially since truncation is also required in accretion onto
black holes (e.g. Dubus, Hameury \& Lasota 2001) where the
magnetic field cannot do the job.

\end{document}